# STATISTICAL SIMULATIONS OF MACHINE ERRORS FOR LINAC4

M. Baylac, JM de Conto, E. Froidefond, LPSC (CNRS/IN2P3-UJF-INPG), Grenoble, France,

E. Sargsyan, CERN, Geneva, Switzerland

*Abstract*

LINAC 4 is a normal conducting H⁻ linac proposed at CERN to provide a higher proton flux to the CERN accelerator chain. It should replace the existing LINAC 2 as injector to the Proton Synchroton Booster and can also operate in the future as the front end of the SPL, a 3.5 GeV Superconducting Proton Linac. LINAC 4 consists of a Radio-Frequency Quadrupole, a chopper line, a Drift Tube Linac (DTL) and a Cell Coupled DTL all operating at 352 MHz and finally a Side Coupled Linac at 704 MHz. Beam dynamics was studied and optimized performing end-to-end simulations. This paper presents statistical simulations of machine errors which were performed in order to validate the proposed design.

## INTRODUCTION

At CERN, the Proton Synchroton Booster is injected by a linear accelerator providing a 50 MeV proton beam with low duty cycle (~0.01%). This injector, LINAC 2, has been operating for nearly 30 years. A new accelerating structure is proposed to increase the proton flux and improve performances of the PS Booster. If realized, this injector, LINAC 4, shall be installed and commissioned by 2010. The proposed structure is a normal conducting linac conceived to accelerate a 65 mA beam of H⁻ ions up to 160 MeV with enhanced duty cycle. Beam dynamics and radio-frequency studies in each section standalone led to the first layout of LINAC 4, which was further optimized via end-to-end simulations [1]. As a final step, statistical simulations with machine tolerances were performed to verify the robustness of this design under realistic conditions.

## LINAC 4

In its initial stage, LINAC 4 will operate as the injector to the PS Booster providing beam at 160 MeV with 0.08% duty cycle. In addition, it is foreseen and designed as the normal conducting front-end of a 3.5 GeV Superconducting Proton Linac (SPL) with an average power 4-5 MW [2]. With such high beam power involved, beam quality must be controlled with extreme care to avoid activation and ensure hands-on maintenance. Even though the SPL duty cycle is 3 to 4%, the machine is entirely designed to allow operation with 15% duty cycle.

The complete layout of LINAC 4 is described in these proceedings in [1]. LINAC 4 starts with a RF source, generating an H⁻ beam at 95 keV. The acceleration is performed up to 3 MeV by a Radio-Frequency Quadrupole resonating at 352 MHz. A chopper is placed at 3 MeV to remove micro-bunches on the RF scale and rematch the beam to the rest of the machine. Two Drift Tube Linacs, conventional Alvarez and Cell Coupled DTL, further boost the beam up to 90 MeV at 352 MHz. The DTL, structured in 3 tanks, brings the beam to 40 MeV. It is fed by 5 klystrons. Beam focusing is performed in the 82 DTL cells with a Permanent Magnet Quadrupoles in each cell. The CCDTL consists of 72 cells powered by 8 klystrons. Electromagnetic Quadrupoles between the 24 tanks provide focusing. The final boost to 160 MeV is achieved via a Side Coupled Linac operating at 704 MHz. The SCL is made of 220 cells and is powered by 4 klystrons. It is equipped with 20 Electromagnetic Quadrupoles.

## MODELLING MACHINE ERRORS

The purpose of this work is to examine the robustness of the LINAC 4 design under realistic conditions and define manufacturing tolerances of the machine components. Tolerances on the Radio-Frequency Quadrupole (IPHI RFQ) have already been decided upon and the RFQ is currently being built. The next system to be manufactured is the Drift Tube Linac, for which tolerances must be established. The error study is performed on the LINAC 4 section consisting of the DTL (13.4 m), the CCDTL (25.2 m) and the SCL (34.5 m).

We model errors on the quadrupole alignment (rotations $\phi_x$, $\phi_y$, $\phi_z$ and transverse translations $\delta_x$, $\delta_y$) and gradient ($\Delta G/G$). Longitudinal effects are also considered by simulating errors on the gap field ($\Delta E_{gap}/E_{gap}$) and on the klystron field and phase ($\Delta E_{klys}/E_{klys}$, $\phi_{klys}$). Each error is applied on all linac cells. For each cell, the amplitude of the error is generated randomly and uniformly within a range of given amplitude. The transport code TraceWin [3] is used for all simulations. At the entrance of the DTL, the beam has an energy of 3 MeV and its normalized RMS emittance is estimated to be $\varepsilon_x = \varepsilon_y = 0.28$ π.mm.mrad and $\varepsilon_z = 0.43$ π.mm.mrad, accounting for the RFQ output including errors.

After chopping, the average beam current is 40 mA, but the average current over the RF pulse is 65 mA. Space charge effects must therefore be determined at 65 mA which is the intensity used for the error study simulations. Space charge interaction is calculated via the 3 dimensional PICNIC routine [4] with a 7x7 mesh, which is a good compromise between accuracy and calculation time. A Gaussian distribution with $5.10^4$ macro-particles per bunch is modelled in the first stage of this work. This number is increased to $10^6$ particles per bunch for the global simulations. This error study is performed in two stages. First, the sensitivity of the structure to one single error is determined in order to evaluate the individual contribution and fix an acceptable limit on each type of error. Then, all errors are combined simultaneously to verify the set of tolerances determined previously and estimate the overall degradation of the beam properties. No correction scheme has been implemented.

## STATISTICAL SIMULATIONS

Each simulation consists of 1000 runs. Beam loss and emittance growth are statistically averaged over the 1000 runs. The relative emittance increase $\Delta\varepsilon$, in each run is expressed with respect to the nominal case, *ie* the case where beam is transported through the ideal linac without errors:

$$\Delta\varepsilon = \frac{\delta\varepsilon_{err} - \delta\varepsilon_{nom}}{\delta\varepsilon_{nom}}$$

where $\delta\varepsilon_{err}$ and $\delta\varepsilon_{nom}$ are the emittance growth of the beam through the structure with and without errors. The natural transverse emittance growth in the nominal case $\delta\varepsilon_{nom}$ is ~9%. As an example, figure 1 displays the statistical distribution of the calculated horizontal emittance increase with respect to the nominal case when all quadrupoles of the linac are rotated around the beam axis by a random angle within [-0.2 deg; +0.2 deg]. For each of the nine types of errors defined above, we perform simulations while varying the maximum allowed amplitude of the error. This aims to determine the amplitude of each error minimizing emittance increase and inducing no beam loss. Discussions with RF and alignment experts along the way ensure that the tolerances thus obtained are achievable

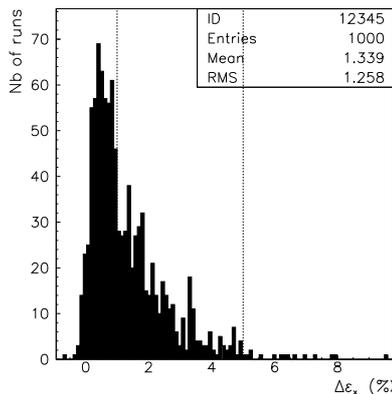

Figure 1: horizontal emittance increase when linac quadrupoles are randomly oriented around the beam direction within ±0.2 deg.

## DTL TOLERANCES

After performing these individual sensitivity simulations, we have determined independently what seems to be an acceptable upper bound for each type of error in the DTL. As an example, figure 2 displays the average emittance increase with respect to the nominal case, if a random roll angle of varying maximum amplitude is applied to all the quadrupoles of the DTL. In this case, the generated emittance growth, similar along both transverse directions, rises quadratically with the roll angle. This behaviour is confirmed by independent theoretical calculations [5]. The tolerance on the roll angle is set to ± 0.2 deg. Then we verify the validity the proposed tolerances and estimate the total degradation of the beam properties using a global error simulation. This lengthy simulation (up to 400 CPU hours) with $10^6$ macro-particles per bunch combines all types of errors simultaneously. The emittance increases in each direction by $\Delta\varepsilon$~4% on average with respect to the nominal case when all errors applied. The transverse beam emittance increase remains below 5% for ~73% for the simulations. No particle loss is detected. We modified the input distribution and verified that the distribution of the emittance increase is simply shifted up or down by ~40% when modelling a Gaussian or a KV distribution. Thus we established the manufacturing tolerances for the DTL to:
- Transverse displacements: $\delta_{x,y} = \pm 0.1$ mm
- Transverse rotations : $\phi_{x,y} = \pm 0.5$ deg
- Longitudinal rotations : $\phi_z = \pm 0.2$ deg
- Gradient: $\Delta G/G = = \pm 0.5$ %,

and for the accelerating field:
- Gap field: $\Delta E_{gap}/ E_{gap} = \pm 1$%
- Klystron field $\Delta E_{klys}/ E_{klys} = \pm 1$%
- Klystron phase $\phi_{klys} = \pm 1$ deg.

These are comparable to the tolerances on other components of LINAC 4 (IPHI RFQ) or other accelerators (SNS). They were accepted by the manufacturer (ITEP-VNIIEF) and by CERN RF experts. The first DTL tank is presently under construction.

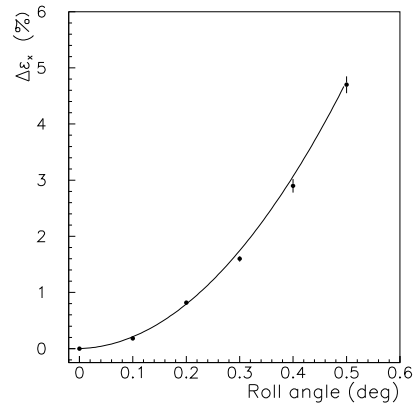

Figure 2: emittance growth when longitudinal rotations are applied to all DTL quadrupoles as a function of the maximum rotation amplitude (plotted for 45 mA). Superposed is a quadratic fit.

## LINAC 4 ERROR STUDY

After the DTL tolerances are decided, we examine the linac section between 3 MeV and 160 MeV. For all transverse and longitudinal effects, errors are applied on the DTL, CCDTL and SCL components. Table 1 presents the average and RMS of the relative emittance growth with respect to the nominal case, as well as the probability for $\Delta\varepsilon$ to be less than 1% or less than 5% in each simulation. Results are symmetric in x and y, such that for example, $\Delta\varepsilon_y \sim 4\%$ for $\delta_y \pm 0.1$ mm. No loss is detected within the quoted amplitudes.

Table 1: sensitivities of the linac to errors

| Error type, amplitude | $\langle\Delta\varepsilon_x\rangle\pm$RMS probabilities | $\langle\Delta\varepsilon_y\rangle\pm$RMS probabilities | $\langle\Delta\varepsilon_z\rangle\pm$ RMS probabilities |
|---|---|---|---|
| $\delta_x$ ± 0.1 mm | 4.1 ± 3.1<br>< 1%: 10.1<br>< 5%: 70.2 | 1.1 ± 0.6<br>< 1%: 55.6<br>< 5%: 99.8 | 3.9 ± 2.7<br>< 1%: 7.8<br>< 5%: 73.9 |
| $\phi_x$ ± 0.5 deg | 0.0 ± 0.1<br>< 1%: 100<br>< 5%: 100 | 0.0 ± 0.1<br>< 1%: 100<br>< 5%: 100 | 0.0 ± 0.1<br>< 1%: 100<br>< 5%: 100 |
| $\phi_z$ ± 0.2 deg | 1.3 ± 1.3<br>< 1%: 53.1<br>< 5%: 98.4 | 1.7 ± 1.0<br>< 1%: 22.9<br>< 5%: 98.7 | 0.1 ± 0.1<br>< 1%: 100<br>< 5%: 100 |
| $\Delta G/G$ ± 0.5% | 0.5 ± 0.7<br>< 1%: 81.8<br>< 5%: 99.8 | 1.2 ± 1.0<br>< 1%: 49.5<br>< 5%: 99.0 | 0.1 ± 0.2<br>< 1%: 99.8<br>< 5%: 100 |
| $\Delta E_{gap}/E_{gap}$ ± 1% | 0.4 ± 0.7<br>< 1%: 79.8<br>< 5%: 99.9 | 0.6 ± 1.1<br>< 1%: 68.4<br>< 5%: 99.3 | 0.5 ± 1.3<br>< 1%: 67.4<br>< 5%: 99.7 |
| $\Delta E_{klys}/E_{klys}$ ± 1% | 1.9 ± 2.0<br>< 1%: 39.6<br>< 5%: 92.4 | 2.3 ± 2.8<br>< 1%: 43.3<br>< 5%: 84.4 | 3.5 ± 5.0<br>< 1%: 32.1<br>< 5%: 75.4 |
| $\phi_{klys}$ ± 1 deg | 1.4 ± 1.4<br>< 1%: 43.9<br>< 5%: 97.6 | 1.8 ± 2.0<br>< 1%: 41.2<br>< 5%: 91.9 | 3.0 ± 3.6<br>< 1%: 31.9<br>< 5%: 78.7 |

Finally, global error simulations are run on the linac. Table 2 summarizes the results obtained when applying the nine errors within the DTL tolerances on the DTL, the CCDTL and the SCL. The sensitive parameters appear to be the quadrupole transverse alignment and longitudinal rotation. Moderate emittance increase is induced by klystron errors or errors on the quadrupole focusing gradient. Very little effect is due to errors on the accelerating field in the gaps or due to transverse quadrupole rotations. We see that the individual sensitivities roughly add up when combining different errors. This observation is useful as one can get a rough estimate of the overall beam degradation using sensitivity runs only, thus avoiding the lengthy global simulations. Under these conditions which account for a realistic linac structure, an average transverse emittance growth with respect to the nominal case is found to be on the order of 15% (see Table 2). In 18 out the 1000 runs, particles are lost along the linac. The estimated power lost is $\sim 0.06$ W/m along the 75 m of the DTL-CCDTL- SCL for a 15% duty cycle. This is well below the acceptable limit of 1 W/m even for such a duty cycle.

Table 2: global error simulations of the linac

| $\langle\Delta\varepsilon_x\rangle\pm$RMS probabilities | $\langle\Delta\varepsilon_y\rangle\pm$RMS probabilities | $\langle\Delta\varepsilon_z\rangle\pm$ RMS probabilities | Lossy runs |
|---|---|---|---|
| 11.3 ± 5.1<br>< 5%: 6.1<br>< 15%: 79.9<br>< 30%: 99.2 | 13.3 ± 6.5<br>< 5%: 2.4<br>< 15%: 69.8<br>< 30%: 98.4 | 18.3 ± 11.9<br>< 5%: 4.1<br>< 15%: 46.9<br>< 30%: 90.0 | 18 out of 1000 |

## SUMMARY

Statistical simulations modelling machine errors were performed on the proposed CERN LINAC 4. This led to the determination of the manufacturing and RF tolerances for the Drift Tube Linac. The DTL is currently under construction. Global simulations were run on the linac section ranging from 3 MeV to 160 MeV to verify the robustness of the design. The beam quality was found to remain good: the emittance growth for all errors uncorrected is ~15% on average. We estimate a particle loss level along the linac well below our acceptable limit, even for operation at 15% duty cycle.

## ACKNOWLEDGEMENTS

We acknowledge the support of the European Community-Research Infrastructure Activity under the FP6 "Structuring the European Research Area" program (CARE, Contract No. RII3-CT-2003-506395).